\documentclass[preprint,eqsecnum,aps,keywords,showpacs]{revtex4}
\usepackage{dcolumn}
\usepackage{graphicx}
\usepackage{amsmath}
\usepackage{amssymb}
\usepackage{epsfig}
\usepackage{float}
\usepackage{latexsym}
\usepackage{rotating}
\usepackage{amsmath}
\usepackage{bm}

\begin{document}

\title{Topological Dyons}
\author{
Vinod Singh\footnote{Electronic address: {\em vinodsingh\_phy@yahoo.com}}${}^{}$}
\affiliation{Department of Physics, Government PG College, Gopeshwar 246 401, India}

\author{
Buddhi Vallabh Tripathi\footnote{Electronic address: {\em buddtrip@rediffmail.com}}${}^{}$}
\affiliation{Department of Physics, APB Government PG College, Agastyamuni 246 421, India}

\begin{abstract}
Using a two potential approach, dyon solutions have been found in the temporal and non-temporal gauges for a non-Abelian theory. Both the charges, electric and magnetic, of the temporal dyon solution are topological, while for the non-temporal case both charges are partially topological.
\\
\\
\hspace{-0.5cm}Keywords: Monopole, Dyon, Two-potential formalism.
\end{abstract}

\maketitle

\section{Introduction}
The works of 't~Hooft \cite{GTH} and Polyakov \cite{POL} demonstrated that a non-Abelian gauge field theory coupled to a self-interacting scalar field gives rise to magnetic monopole. This magnetic monopole is stable because of the nontrivial topology of the vacuum expectation value of the scalar field. Extending the discussion of  't~Hooft \cite{GTH}, Julia-Zee \cite{JZEE} showed that a non-Abelian gauge theory with Higgs fields exhibits classical solution having both electrical and magnetic  charges, which are called dyons.  However the electric charge cannot exist without an accompanying magnetic charge and the stability arguments that apply to the dyonic solution. The conventional theories of monopoles and dyons show the presence of singularities. Dual potential approach has been explored to give singularity free formalism for the theories of monopoles and dyons. Cabibbo and Ferrari \cite{CABFER} gave a two potential theory  for Abelian dyons.  In this connection Zwanzinger \cite{ZWA1,ZWA2} developed the quantum field theory of particles with electric and magnetic charges as an extension of Schwinger's \cite{SCH1,SCH2} quantum field theory  of particles with either electric or magnetic charge. Benjwal and Joshi \cite{BENJOS} extended the Cabibbo-Ferrari \cite{CABFER} approach to the non-Abelian case by employing a non-Abelian field tensor. The same non-Abelian field tensor has also been used to obtain dyon solutions for temporal and non-temporal cases for a $SU(3)\otimes SU(3)$ gauge theory \cite{SVINOD,SVINOD1}.

Singleton \cite{STON} has given a (symmetric) formulation of electrodynamics which employs two four-vector potentials and the magnetic charge appears as a gauge charge without any singularities. In the present paper we extend Singleton's approach to a non-Abelian theory and derive dyon solutions for the temporal and non-temporal gauges. The distinguishing feature of the obtained solutions is the topological origin of both electric and magnetic charges.

Section~II describes the Lagrangian and field equations. Particular finite energy temporal solutions for $SU(2)$ have been derived in Section~III. The non-Abelian theory has been reduced to the Abelian Singleton theory in Section~IV and thereby electrical and magnetic charges have been obtained. In Section~V the mass of the obtained dyon solutions has been derived and non-temporal dyon solutions are presented in Section~VI.

\section{The Lagrangian and Field Equations}

The Lagrangian density that we consider is \cite{SVINOD}
\begin{eqnarray}\label{lagden2}
\mathcal{L}=-\frac{1}{4}A_{\mu\nu}^a A^{\mu\nu a}+\frac{1}{4}\widetilde B_{\mu\nu}^a \widetilde B^{\mu\nu a} +\frac{1}{2}(D_\mu^1\phi_e^a)(D^{1\mu}\phi_e^a)
+\frac{1}{2}(D_\mu^2\phi_g^a)(D^{2\mu}\phi_g^a)+V(\phi_e^a,\phi_g^a)
\end{eqnarray}
where
\begin{eqnarray}\label{defs}
&A&_{\mu\nu}^a=\partial_\mu A_\nu^a-\partial_\nu A_\mu^a-ef^{abc}A_\mu^b A_\nu^c,\qquad\qquad D^1_\mu\phi_e^a=\partial_\mu\phi_e^a-e f^{abc}A_\mu^b\phi_e^c,\nonumber\\
&B&_{\mu\nu}^a=\partial_\mu B_\nu^a-\partial_\nu B_\mu^a-g f^{abc}B_\mu^b B_\nu^c, \qquad\qquad D^2_\mu\phi_g^a=\partial_\mu\phi_g^a-g f^{abc}B_\mu^b\phi_g^c,\nonumber\\
&\widetilde B&_{\mu\nu}^a=\frac{1}{2}\varepsilon_{\mu\nu\rho\sigma}B^{\rho\sigma a}.
\end{eqnarray}
The potential energy has the form $V(\phi_e^a,\phi_g^a)=-\eta(\phi_e^a\phi_e^a+\phi_g^a\phi_g^a-\xi^2)^2$ where $\eta$ and $\xi$ are real constants with $\eta\geq 0$. The gauge group under consideration is $SU(2)$, for which the structure constants $f^{abc}=\varepsilon^{abc}$.
The Lagrangian density (\ref{lagden2}) is invariant under the following two independent sets of gauge transformations
\begin{eqnarray}
\phi_e\rightarrow U_1\phi_e,\qquad\qquad A_\mu\rightarrow U_1A_\mu U_1^{-1}+\frac{\iota}{e}(\partial_\mu U_1)U_1^{-1},\nonumber\\
\phi_g\rightarrow U_2\phi_g,\qquad\qquad B_\mu\rightarrow U_2B_\mu U_2^{-1}+\frac{\iota}{g}(\partial_\mu U_2)U_2^{-1},
\end{eqnarray}
where,
\begin{eqnarray}
\phi_e=\phi_e^a T^a,\qquad\qquad A_\mu=A_\mu^a T^a,\qquad\qquad U_1={exp}(-\iota\theta_1^a T^a),\nonumber\\
\phi_g=\phi_g^a T^a,\qquad\qquad B_\mu=B_\mu^a T^a,\qquad\qquad U_2={exp}(-\iota\theta_2^a T^a),
\end{eqnarray}
with $T^a$ being the generators of the gauge group ($SU(2)$ for this case) and $\theta$'s are some space-time parameters.

The Lagrangian density (\ref{lagden2}) gives the field equations
\begin{eqnarray}\label{feqn1}
D^1_\mu A^{\mu\nu a}-e f^{abc} \phi_e^b D^{1\nu}\phi_e^c&=&0,\nonumber\\
D^2_\mu B^{\mu\nu a}-g f^{abc} \phi_g^b D^{2\nu}\phi_g^c&=&0,\nonumber\\
D^1_\mu(D^{1\mu}\phi_e^a)+4\eta\phi_e^a(\phi_e^b\phi_e^b+\phi_g^b\phi_g^b-\xi^2)&=&0,\nonumber\\
D^2_\mu(D^{2\mu}\phi_g^a)+4\eta\phi_g^a(\phi_e^b\phi_e^b+\phi_g^b\phi_g^b-\xi^2)&=&0.
\end{eqnarray}
We assume that all fields $(A_\mu^a, B_\mu^a,\phi_e^a,\phi_g^a)$ are static i.e. time-independent. We also assume the temporal gauge condition for the fields $A_\mu^a$ and $B_\mu^a$, i.e.
\begin{equation}\label{tgauge}
A_\mu^0(\bm{r})=0=B_\mu^0(\bm{r}).
\end{equation}
Under these assumptions the field equations (\ref{feqn1}) reduce to
\begin{eqnarray}\label{feqn2}
D^1_i A^{ija}-e f^{abc} \phi_e^b D^{1j}\phi_e^c&=&0,\nonumber\\
D^2_i B^{ija}-g f^{abc} \phi_g^b D^{2j}\phi_g^c&=&0,\nonumber\\
D^1_i(D^{1i}\phi_e^a)+4\eta\phi_e^a(\phi_e^b\phi_e^b+\phi_g^b\phi_g^b-\xi^2)&=&0,\nonumber\\
D^2_i(D^{2i}\phi_g^a)+4\eta\phi_g^a(\phi_e^b\phi_e^b+\phi_g^b\phi_g^b-\xi^2)&=&0
\end{eqnarray}
which are equations of motion of static sourceless Euclidean Yang-Mills fields $A_\mu^a$ and $B_\mu^a$.

\section{Finite Energy Solutions}
We closely follow Rajaraman (Section 3.4) \cite{RRN} to derive a particular finite energy solution for this system. The Hamiltonian density of the system is
\begin{eqnarray}\label{hamden}
\mathcal{H}=\frac{1}{4}A_{ij}^a A^{ija}&-&\frac{1}{4}(\widetilde B_{0i}^a \widetilde B^{0ia}+\widetilde B_{i0}^a \widetilde B^{i0a})-\frac{1}{2}(D^1_i\phi_e^a)(D^{1i}\phi_e^a)\nonumber\\*
&-&\frac{1}{2}(D^2_i\phi_g^a)(D^{2i}\phi_g^a)+\eta(\phi_e^a\phi_e^a+\phi_g^a\phi_g^a-\xi^2)^2\nonumber\\
=\frac{1}{4}(A_{ij}^a)^2&+&\frac{1}{4}(B_{ij}^a)^2+\frac{1}{2}(D^1_i\phi_e^a)^2\nonumber\\*
&+&\frac{1}{2}(D^2_i\phi_g^a)^2+\eta(\phi_e^a\phi_e^a+\phi_g^a\phi_g^a-\xi^2)^2.
\end{eqnarray}
The total energy of the system given by $E=\int d^3 x \mathcal{H}$ is clearly $\geq 0$. The energy vanishes (minimum) for \cite{RRN}
\begin{eqnarray}\label{fcond}
A_i^a=0=B_i^a,\quad\qquad D^1_i\phi_e^a=0=D^2_i\phi_g^a,\quad\qquad (\phi_e^a)^2+(\phi_g^a)^2=\xi^2.
\end{eqnarray}
The system will have a finite energy only if the fields achieve the energy vanishing conditions of Eq.~(\ref{fcond}) at spatial infinity sufficiently fast, i.e.
\begin{eqnarray}\label{bcond}
r^{\frac{3}{2}}D^1_i\phi_e^a\rightarrow 0,\quad r^{\frac{3}{2}}D^2_i\phi_g^a\rightarrow 0,\quad (\phi_e^a)^2+(\phi_g^a)^2\rightarrow\xi^2 \qquad\text{as}\qquad r\rightarrow \infty.
\end{eqnarray}
A particular finite energy solution can be obtained by employing the 't Hooft-Ployakov \cite{GTH,POL} (temporal) ansatz for the fields
\begin{eqnarray}\label{ansz}
\phi_e^a=\delta_i^a\frac{x^i}{r}F_1(r),\qquad\qquad &&A_i^a=\varepsilon_{aij}\frac{x^j}{r}W_1(r),\nonumber\\*
\phi_g^a=\delta_i^a\frac{x^i}{r}F_2(r),\qquad\qquad &&B_i^a=\varepsilon_{aij}\frac{x^j}{r}W_2(r).
\end{eqnarray}
The boundary conditions (\ref{bcond}) for finite-energy configuration then imply \cite{RRN}
\begin{eqnarray}\label{const}
W_1(r)\rightarrow\frac{1}{er},\quad\quad W_2(r)\rightarrow\frac{1}{gr},\quad\quad\{F_1(r)\}^2+\{F_2(r)\}^2\rightarrow\xi^2\quad{as}\quad r\rightarrow \infty.
\end{eqnarray}
Plugging the ansatz (\ref{ansz}) into the field equations (\ref{feqn2}), we get for the $\eta=0$ case 
\begin{eqnarray}\label{feqn3}
r^2K_1^{''}=K_1({K_1}^2+{H_1}^2-1),\qquad\qquad&& r^2H_1^{''}=2H_1{K_1}^2,\nonumber\\*
r^2K_2^{''}=K_2({K_2}^2+{H_2}^2-1),\qquad\qquad&& r^2H_2^{''}=2H_2{K_2}^2,
\end{eqnarray}
where
\begin{eqnarray}\label{kwhf}
K_1(r)=1-erW_1(r),\qquad\qquad\qquad && H_1(r)=erF_1(r),\nonumber\\*
K_2(r)=1-grW_2(r),\qquad\qquad\qquad && H_2(r)=grF_2(r),
\end{eqnarray}
and $K_1^{''}$ denotes $\frac{d^2}{dr^2}K_1(r)$ and so on.

The solution of Eqs. (\ref{feqn3}), satisfying the constraints of finite energy (\ref{const}) are \cite{RRN,PSF}
\begin{eqnarray}\label{parsoln}
K_1=\frac{erC}{\sinh (erC)},\qquad\qquad\qquad\qquad H_1=\frac{erC}{\tanh (erC)}-1,\qquad\nonumber\\
K_2=\frac{gr\sqrt{\xi^2-C^2}}{\sinh (gr\sqrt{\xi^2-C^2})},\qquad\qquad H_2=\frac{gr\sqrt{\xi^2-C^2}}{\tanh (gr\sqrt{\xi^2-C^2})}-1,
\end{eqnarray}
where $C$ is some constant such that $0\leq C\leq \xi$.

\section{Topological Electric and Magnetic Charges}
Maxwell's equations for electrodynamics can be symmetrized by introducing magnetic charge and current
\begin{eqnarray}
\bm{\nabla}\cdot\bm{E}=\rho_e,\qquad\qquad&&\bm{\nabla}\times\bm{B}=\frac{\partial\bm{E}}{\partial t}+\bm{J}_e,\nonumber\\*
\bm{\nabla}\cdot\bm{B}=\rho_m,\qquad\qquad&-&\bm{\nabla}\times\bm{E}=\frac{\partial\bm{B}}{\partial t}+\bm{J}_m.
\end{eqnarray}
The $\bm{E}$ and $\bm{B}$ field vectors are distinguished from one another under the spatial inversion or parity transformation $(\bm{r}\rightarrow-\bm{r})$, $\bm{E}$ being a vector $(\bm{E}\rightarrow-\bm{E})$ and $\bm{B}$ being a pseudovector $(\bm{B}\rightarrow\bm{B})$. These definitions and Maxwell's equations then imply that $\bm{J}_e$ and $\bm{J}_m$ are vector and pseudovector, respectively under parity while $\rho_e$ and $\rho_m$ are scalar and pseudoscalar, respectively under parity.

Singleton \cite{STON} has given a formulation of electrodynamics with electric and magnetic charges which employs two four-potentials. In this formalism the electric and magnetic fields are given by
\begin{equation}\label{emfields}
E_i=F^{i0}-\widetilde G^{i0},\qquad\qquad\qquad B_i=G^{i0}+\widetilde F^{i0},
\end{equation}
where the field tensors and their duals are defined in terms of the two potentials $A^\mu$ and $B^\mu$ as
\begin{eqnarray}\label{sinten}
F_{\mu\nu}=\partial_\mu A_\nu-\partial_\nu A_\mu,\qquad\qquad&& G_{\mu\nu}=\partial_\mu B_\nu-\partial_\nu B_\mu,\nonumber\\*
\widetilde F^{\mu\nu}=\frac{1}{2}\varepsilon^{\mu\nu\rho\sigma}F_{\rho\sigma},\qquad\qquad&&\widetilde G^{\mu\nu}=\frac{1}{2}\varepsilon^{\mu\nu\rho\sigma}G_{\rho\sigma}.
\end{eqnarray}
Here $\varepsilon^{\mu\nu\rho\sigma}$ is the completely antisymmetric tensor in four dimensions with the choice $\varepsilon^{0123}=+1$.

Following 't Hooft \cite{GTH}, we define two gauge invariant field tensors
\begin{eqnarray}\label{hoften}
\mathcal{F}_{\mu\nu}&=&\widehat\phi_e^a\ A_{\mu\nu}^a -\frac{1}{e}\ \varepsilon^{abc}\ \widehat\phi_e^a\ D^1_\mu \widehat\phi_e^b\ D^1_\nu \widehat\phi_e^c,\nonumber\\*
\mathcal{G}_{\mu\nu}&=&\widehat\phi_g^a\ B_{\mu\nu}^a -\frac{1}{g}\ \varepsilon^{abc}\ \widehat\phi_g^a\ D^2_\mu \widehat\phi_g^b\ D^2_\nu \widehat\phi_g^c,
\end{eqnarray}
where $\widehat\phi_e^a=\phi_e^a/\sqrt{\phi_e^a\phi_e^a}$ and $\widehat\phi_g^a=\phi_g^a/\sqrt{\phi_g^a\phi_g^a}$. We adopt the signature (-,+,+,+) so that $\varepsilon^{123}=\varepsilon_{123}=+1$. For a gauge in which $\widehat\phi_e^a$ is constant, say in the three-direction, the tensor $\mathcal{F}_{\mu\nu}$ reduces to \cite{GTH,WBERG} $\mathcal{F}_{\mu\nu}=\partial_\mu A_\nu^3-\partial_\nu A_\mu^3$, which is of the form of $F_{\mu\nu}$ (See Eq.~(\ref{sinten})). Similarly in a gauge with constant $\widehat\phi_g^a$, $\mathcal{G}_{\mu\nu}$ becomes similar to $G_{\mu\nu}$. Thus the 't Hooft tensors (\ref{hoften}) reduce to the Singleton tensors (\ref{sinten}) for particular gauge choices.

We now employ $\mathcal{F}_{\mu\nu}$ and $\mathcal{G}_{\mu\nu}$ instead of $F_{\mu\nu}$ and $G_{\mu\nu}$ in Eq. (\ref{emfields}) to derive the electric and magnetic fields
\begin{equation}\label{emfields2}
E_i=\mathcal{F}^{i0}-\widetilde{\mathcal{G}}^{i0},\qquad\qquad\qquad B_i=\mathcal{G}^{i0}+\widetilde{\mathcal{F}}^{i0}.
\end{equation}
As $\mathcal{F}^{i0}=0=\mathcal{G}^{i0}$ due to static and temporal-gauge choice for the fields we have
\begin{eqnarray}\label{emfields1}
&&E_i=-\widetilde{\mathcal{G}}^{i0}=-\frac{1}{2}\varepsilon^{i0\rho\sigma}\mathcal{G}_{\rho\sigma}=\frac{1}{2}\varepsilon^{ijk}\mathcal{G}_{jk},\nonumber\\*
&&B_i=\widetilde{\mathcal{F}}^{i0}=\frac{1}{2}\varepsilon^{i0\rho\sigma}\mathcal{F}_{\rho\sigma}=-\frac{1}{2}\varepsilon^{ijk}\mathcal{F}_{jk}.
\end{eqnarray}

We also define two currents \cite{AFG}
\begin{eqnarray}\label{crnts}
(k_e)_\mu&=&\frac{1}{8\pi}\ \varepsilon_{\mu\nu\rho\sigma}\ \varepsilon_{abc}\ \partial^\nu\widehat\phi_e^a\ \partial^\rho\widehat\phi_e^b\ \partial^\sigma\widehat\phi_e^c,\nonumber\\*
(k_g)_\mu&=&\frac{1}{8\pi}\ \varepsilon_{\mu\nu\rho\sigma}\ \varepsilon_{abc}\ \partial^\nu\widehat\phi_g^a\ \partial^\rho\widehat\phi_g^b\ \partial^\sigma\widehat\phi_g^c.
\end{eqnarray}
Both these currents are conserved by construction itself i.e. ${\partial^\mu(k_e)_\mu= 0 =\partial^\mu(k_g)_\mu}$. The corresponding conserved (topological) charges are \cite{AFG}
\begin{equation}\label{topcrg}
Q_1=\int d^3x\ (k_e)^0,\qquad\qquad\qquad Q_2=\int d^3x\ (k_g)^0.
\end{equation}
These charges can be interpreted as the topological winding numbers of the corresponding $\phi$-fields. The currents~(\ref{crnts}) are related to the 't~Hooft tensors (\ref{hoften}) as
\begin{eqnarray}\label{crten}
(k_e)_\mu=\frac{e}{8\pi}\varepsilon_{\mu\nu\rho\sigma}\partial^\nu\mathcal{F}^{\rho\sigma},\qquad\qquad\qquad
(k_g)_\mu=\frac{g}{8\pi}\varepsilon_{\mu\nu\rho\sigma}\partial^\nu\mathcal{G}^{\rho\sigma}.
\end{eqnarray}

The magnetic flux can be computed using Eqs. (\ref{emfields1}), (\ref{crnts}) and (\ref{crten}) as
\begin{eqnarray}
\bm{\nabla}\cdot\bm{B}=\partial^i B_i=\partial^i(-\frac{1}{2}\varepsilon^{ijk}\mathcal{F}_{jk})=\frac{1}{2}\varepsilon_{0\nu\rho\sigma}(\partial^\nu\mathcal{F}^{\rho\sigma})=-\frac{4\pi}{e}(k_e)^0.
\end{eqnarray}
The magnetic charge given by the volume integral of the magnetic flux is
\begin{eqnarray}\label{magchrg}
q_m=\int d^3x\ (\bm{\nabla}\cdot\bm{B})=\int d^3x\ (-\frac{4\pi}{e})(k_e)^0=-\frac{4\pi}{e}Q_1,
\end{eqnarray}
where $Q_1$ is the topological charge (See Eq. (\ref{topcrg})). Similarly the electric charge is
\begin{equation}\label{elecchrg}
q_e=\frac{4\pi}{g}Q_2.
\end{equation}

Both these charges, derived for the temporal gauge choice, are topological in origin. The particular solutions~(\ref{parsoln}), derived above by employing the ansatz (\ref{ansz}), correspond to $Q_1=1=Q_2$ and thus are dyon-solutions with electric and magnetic charges $4\pi/g$ and $-4\pi/e$ respectively.

\section{Mass of the Dyon}
A Bogomol'nyi \cite{BOG} type bound can be derived for the dyon-solutions derived above. The energy of the system~(\ref{hamden}) for the $\eta=0$ case can be written as
\begin{eqnarray}\label{eng}
E=\int d^3x\bigg{[}\sum_{i,j,a}\bigg\{\frac{1}{4}(A_{ij}^a-\varepsilon_{ijk}D^1_k\phi_e^a)^2&+&\frac{1}{4}(B_{ij}^a-\varepsilon_{ijk}D^2_k\phi_g^a)^2\bigg\}\nonumber\\
&+&\frac{1}{2}\varepsilon_{ijk}A_{ij}^aD^1_k\phi_e^a+\frac{1}{2}\varepsilon_{ijk}B_{ij}^aD^2_k\phi_g^a\bigg{]}.
\end{eqnarray}
The third and fourth terms in the above integral (\ref{eng}) can be written as surface integrals of the form
\begin{equation}\label{sfcint}
\oint d\sigma_k \big{(}\frac{1}{2}\varepsilon_{kij}A_{ij}^a\phi_e^a\big{)}
\quad\qquad\text{and}\qquad\quad
\oint d\sigma_k\big{(}\frac{1}{2}\varepsilon_{kij}B_{ij}^a\phi_g^a\big{)},
\end{equation}
where the integration is to be performed at the surface of a two-sphere at spatial infinity. At spatial infinity, $D^1_i\phi_e^a=0$, due to boundary condition  for finite energy (see Eq. (\ref{bcond})). Also, $\hat\phi_e^a=\phi_e^a/C$ at spatial infinity (Using Eqs. (\ref{ansz}), (\ref{kwhf}), (\ref{parsoln})), so that the space-space component of the 't~Hooft tensor $\mathcal{F}_{\mu\nu}$ (\ref{hoften}) is $\mathcal{F}_{ij}=(\phi_e^a/C)A_{ij}^a$. Hence, the magnetic field (\ref{emfields1})(at spatial infinity) can be written as $B_i=-\frac{1}{2}\varepsilon^{ijk}(\phi_e^a/C)A_{jk}^a$. The first surface integral in Eq.~(\ref{sfcint}) can therefore be rewritten as
\begin{eqnarray}\label{sur1}
\oint d\sigma_k\big{(}\frac{1}{2}\varepsilon_{kij}A_{ij}^a\phi_e^a\big{)}=\oint d\sigma_k \big{(}-C B^k\big{)}=-C (q_m)=\frac{4 \pi}{e} C,
\end{eqnarray}
where use of Eq. (\ref{magchrg}), with $Q_1=1$, is implied. The second surface integral in Eq.~(\ref{sfcint}), similarly, simplifies to
\begin{equation}\label{sur2}
\oint d\sigma_k\big{(}\frac{1}{2}\varepsilon_{kij}B_{ij}^a\phi_g^a\big{)}=\frac{4\pi}{g}\sqrt{\xi^2-C^2}.
\end{equation}

Using Eqs. (\ref{sur1}) and (\ref{sur2}), Eq. (\ref{eng}) becomes
\begin{eqnarray}\label{engeq}
E&=&\frac{4\pi}{e}C+\frac{4\pi}{g}\sqrt{\xi^2-C^2}\nonumber\\*
&&\qquad+\int d^3x\sum_{i,j,a}\bigg\{\frac{1}{4}(A_{ij}^a-\varepsilon_{ijk}D^1_k\phi_e^a)^2+\frac{1}{4}(B_{ij}^a-\varepsilon_{ijk}D^2_k\phi_g^a)^2\bigg\}\nonumber\\*
&\geq&\frac{4\pi}{e}C+\frac{4\pi}{g}\sqrt{\xi^2-C^2}.
\end{eqnarray}
The energy reaches its minimum when \cite{BOG}
\begin{eqnarray}
A_{ij}^a&=&\varepsilon_{ijk}D^1_k\phi_e^a,\qquad\qquad\qquad B_{ij}^a=\varepsilon_{ijk}D^2_k\phi_g^a.
\end{eqnarray}
These are the Bogomol'nyi-type conditions for the topological dyon. The equality in Eq. (\ref{engeq}) corresponds to the mass of the dyon solution (\ref{parsoln}). The energy of the monopole configuration in the BPS limit is independent of the properties of the gauge fields and completely defined by the Higgs field alone.

\section{Non-Temporal Solutions}
If instead of the temporal gauge choice (\ref{tgauge}) we use the Julia-Zee ansatz \cite{JZEE} for $A_0^a$ and $B_0^a$ i.e.
\begin{equation}\label{jzansz}
A_0^a=\frac{x^a}{er^2}J_1(r),\qquad\qquad\qquad B_0^a=\frac{x^a}{gr^2}J_2(r)
\end{equation}
along with those in Eq. (\ref{ansz}), the solutions (\ref{parsoln}) are modified as \cite{JZEE}
\begin{eqnarray}\label{parsoln1}
K_1=\frac{erC}{\sinh (erC)},\qquad\qquad\qquad\qquad K_2&=&\frac{gr\sqrt{\xi^2-C^2}}{\sinh(gr\sqrt{\xi^2-C^2})},\nonumber\\
H_1=\cosh\gamma_1\bigg{[}\frac{erC}{\tanh(erC)}-1\bigg{]},\qquad\quad H_2&=&\cosh\gamma_2\bigg{[}\frac{gr\sqrt{\xi^2-C^2}}{\tanh(gr\sqrt{\xi^2-C^2})}-1\bigg{]},\nonumber\\
J_1=\sinh\gamma_1\bigg{[}\frac{erC}{\tanh(erC)}-1\bigg{]},\qquad\quad J_2&=&\sinh\gamma_2\bigg{[}\frac{gr\sqrt{\xi^2-C^2}}{\tanh(gr\sqrt{\xi^2-C^2})}-1\bigg{]},
\end{eqnarray}
where $\gamma_1$ and $\gamma_2$ are arbitrary constants. For the temporal case $\mathcal{F}^{i0}$ and $\mathcal{G}^{i0}$ were zero and the electric and magnetic charges were as given by Eq. (\ref{emfields1}). For the non-temporal case, $\mathcal{F}^{i0}$ and $\mathcal{G}^{i0}$ being non-zero, the electric and magnetic fields are to be computed using Eq.~(\ref{emfields2}). The magnetic charge calculated in Eq. (\ref{magchrg}) now contains an extra contribution $ \int d^3 x (\partial^i\mathcal{G}^{i0})$ which simplifies to $(1/g)\int d^3 x (J_2^{''}/r)$ on using Eqs. (\ref{defs}),
(\ref{ansz}), (\ref{hoften}) and (\ref{jzansz}). This integral when evaluated using Eq. (\ref{parsoln1}) gives $(4\pi/g)\sinh{\gamma_2}$~\cite{PSF}. Therefore, the total magnetic charge of the non-temporal solution (\ref{parsoln1}) is
\begin{equation}
q_m=-\frac{4\pi}{e}+\frac{4\pi}{g}\sinh{\gamma_2}.
\end{equation}
Similarly, the total electric charge of the solution (\ref{parsoln1}) is
\begin{equation}
q_e=\frac{4\pi}{g}+\frac{4\pi}{e}\sinh{\gamma_1}.
\end{equation}
Thus, the non-temporal solutions (\ref{parsoln1}) are also dyonic solutions. There is a possibility of the charge $q_e$ (or $q_m$), of the non-temporal solution (\ref{parsoln1}), being zero for particular value of $\gamma_2$ (or $\gamma_1$), so that we are left with a pure monopole (or pure electric charge) only.

\section{Summary}
The two-potential Cabibbo-Ferrari Abelian theory of magnetic monopole has been extended by a non-Abelian theory. Using the ’t Hooft-Polyakov ansatz, finite energy solutions have been obtained in the temporal gauge. These solutions carry (topological) electric and magnetic charges and thus are dyonic-solutions. The energy of these solutions is found to be bounded from below. The temporal gauge choice when extended to the non-temporal gauge (using the Julia-Zee ansatz) again yields solutions carrying electric and magnetic charges.

In the conventional (dual) theories of magnetic monopoles and dyons the magnetic charge (electric charge) is topological in origin. The distinguishing feature of the solutions obtained here is that both the charges (electric and magnetic) are completely topological in origin, for the temporal case and partially-topological for the non-temporal case. Furthermore, in the conventional theories the dyon solutions emerge only in the non-temporal case whereas in this two-potential formulation the dyon solutions are present even in the temporal case.The energy of the temporal case dyon solutions as derived in Section V is positive definite and bounded from below which guarantees the stability of the obtained solutions. However, a two-potential theory implies presence of two photons when the theory is quantized. Experimentally, only one photon is observed in nature. A possible remedy for this might be (a large) mass generation for one of the photons via symmetry breaking mechanism \cite{STON,STON1}. It is well known that monopoles and dyons emerge as essential ingredients in the dual
superconductor models of QCD in context with the quark confinement problem \cite{RIPKA,GREENSITE,HEM1,HEM2}. It would be interesting to investigate the issue of confinement through the condensation of
these topological dyons.


\begin{thebibliography}{25}

\expandafter\ifx\csname natexlab\endcsname\relax\def\natexlab#1{#1}\fi
\expandafter\ifx\csname bibnamefont\endcsname\relax
  \def\bibnamefont#1{#1}\fi
\expandafter\ifx\csname bibfnamefont\endcsname\relax
  \def\bibfnamefont#1{#1}\fi
\expandafter\ifx\csname citenamefont\endcsname\relax
  \def\citenamefont#1{#1}\fi
\expandafter\ifx\csname url\endcsname\relax
  \def\url#1{\texttt{#1}}\fi
\expandafter\ifx\csname urlprefix\endcsname\relax\def\urlprefix{URL }\fi
\providecommand{\bibinfo}[2]{#2}
\providecommand{\eprint}[2][]{\url{#2}}


\bibitem{GTH}
\bibinfo{author}{\bibfnamefont{'t Hooft}, \bibnamefont{G.}}:
\bibinfo{title}{Magnetic monopoles in unified gauge theories}.
  \bibinfo{journal}{Nucl. Phys. B} \textbf{\bibinfo{volume}{79}},
  \bibinfo{pages}{276-284} (\bibinfo{year}{1974})

\bibitem{POL}
\bibinfo{author}{\bibfnamefont{Polyakov}, \bibnamefont{A.M.}}:
\bibinfo{title}{Particle spectrum in quantum field theory}.
  \bibinfo{journal}{JETP Lett.} \textbf{\bibinfo{volume}{20}},
  \bibinfo{pages}{194-195} (\bibinfo{year}{1974}) [Pis'ma Zh. Eksp. Teor. Fiz. \textbf{20}, 430-433 (1974)]

\bibitem{JZEE}
\bibinfo{author}{\bibfnamefont{Julia}, \bibnamefont{B.}},
 \bibinfo{author}{\bibfnamefont{Zee}, \bibnamefont{A.}}:
\bibinfo{title}{Poles with both magnetic and electric charges in non-Abelian gauge theory}.
  \bibinfo{journal}{Phys. Rev. D} \textbf{\bibinfo{volume}{11}},
  \bibinfo{pages}{2227-2232} (\bibinfo{year}{1975})

\bibitem{CABFER}
\bibinfo{author}{\bibfnamefont{Cabibbo}, \bibnamefont{N.}},
 \bibinfo{author}{\bibfnamefont{Ferrari}, \bibnamefont{E.}}:
\bibinfo{title}{Quantum electrodynamics with Dirac monopole}.
  \bibinfo{journal}{Nuovo Cimento} \textbf{\bibinfo{volume}{23}},
  \bibinfo{pages}{1147-1154} (\bibinfo{year}{1962})

\bibitem{ZWA1}
\bibinfo{author}{\bibfnamefont{Zwanziger}, \bibnamefont{D.}}:
\bibinfo{title}{Exactly soluble nonrelativistic model of particles with both electric and magnetic charges}.
  \bibinfo{journal}{Phys. Rev.} \textbf{\bibinfo{volume}{176}},
  \bibinfo{pages}{1480-1488} (\bibinfo{year}{1968})

\bibitem{ZWA2}
\bibinfo{author}{\bibfnamefont{Zwanziger}, \bibnamefont{D.}}:
\bibinfo{title}{Quantum field theory of particles with both electric and magnetic charges}.
  \bibinfo{journal}{Phys. Rev.} \textbf{\bibinfo{volume}{176}},
  \bibinfo{pages}{1489-1495} (\bibinfo{year}{1968})

\bibitem{SCH1}
\bibinfo{author}{\bibfnamefont{Schwinger}, \bibnamefont{J.}}:
\bibinfo{title}{Magnetic charge and quantum field theory}.
  \bibinfo{journal}{Phys. Rev.} \textbf{\bibinfo{volume}{144}},
  \bibinfo{pages}{1087-1093} (\bibinfo{year}{1966})

\bibitem{SCH2}
\bibinfo{author}{\bibfnamefont{Schwinger}, \bibnamefont{J.}}:
\bibinfo{title}{Electric- and magnetic-charge renormalization. I}.
  \bibinfo{journal}{Phys. Rev.} \textbf{\bibinfo{volume}{151}},
  \bibinfo{pages}{1048-1054} (\bibinfo{year}{1966});
 \bibinfo{title}{Electric- and magnetic-charge renormalization. II}.
  \bibinfo{journal}{Phys. Rev.} \textbf{\bibinfo{volume}{151}},
  \bibinfo{pages}{1055-1057} (\bibinfo{year}{1966})

\bibitem{BENJOS}
\bibinfo{author}{\bibfnamefont{Benjwal}, \bibnamefont{M.P.}},
  \bibinfo{author}{\bibfnamefont{Joshi}, \bibnamefont{D.C.}}:
\bibinfo{title}{Dyon solutions in the temporal gauge}.
  \bibinfo{journal}{Phys. Rev. D} \textbf{\bibinfo{volume}{36}},
  \bibinfo{pages}{629-631} (\bibinfo{year}{1987})

\bibitem{SVINOD}
\bibinfo{author}{\bibfnamefont{Singh}, \bibnamefont{V.}},
  \bibinfo{author}{\bibfnamefont{Tripathi}, \bibnamefont{B.V.}},
  \bibinfo{author}{\bibfnamefont{Joshi}, \bibnamefont{D.C.}}:
\bibinfo{title}{Euclidean space dyon solutions}.
  \bibinfo{journal}{Indian J. Pure \& Appl. Phys.}
  \textbf{\bibinfo{volume}{43}}, \bibinfo{pages}{157-166} (\bibinfo{year}{2005})

\bibitem{SVINOD1}
\bibinfo{author}{\bibfnamefont{Singh}, \bibnamefont{V.}},
  \bibinfo{author}{\bibfnamefont{Tripathi}, \bibnamefont{B.V.}},
  \bibinfo{author}{\bibfnamefont{Joshi}, \bibnamefont{D.C.}}:
\bibinfo{title}{Stability analysis of dyon solutions in SU(3)~$\otimes$~SU(3) gauge theory}.
  \bibinfo{journal}{Indian J. Pure \& Appl. Phys.}
  \textbf{\bibinfo{volume}{44}}, \bibinfo{pages}{567-575} (\bibinfo{year}{2006})

\bibitem{STON}
\bibinfo{author}{\bibfnamefont{Singleton}, \bibnamefont{D.}}:
\bibinfo{title}{Electromagnetism with magnetic charge and two photons}.
  \bibinfo{journal}{Am. J. Phys.} \textbf{\bibinfo{volume}{64}},
  \bibinfo{pages}{452-458} (\bibinfo{year}{1996})

\bibitem{RRN}
\bibinfo{author}{\bibfnamefont{Rajaraman}, \bibnamefont{R.}}:
  \emph{\bibinfo{title}{Solitons and Instantons}}
  (\bibinfo{publisher}{North-Holland Publishing Company},
  \bibinfo{year}{1982})

\bibitem{PSF}
\bibinfo{author}{\bibfnamefont{Prasad}, \bibnamefont{M.K.}},
  \bibinfo{author}{\bibfnamefont{Sommerfield}, \bibnamefont{C.M.}}:
\bibinfo{title}{Exact classical solution for the 't~Hooft monopole and the Julia-Zee dyon}.
  \bibinfo{journal}{Phys. Rev. Lett.} \textbf{\bibinfo{volume}{35}},
  \bibinfo{pages}{760-762} (\bibinfo{year}{1975})

\bibitem{WBERG}
\bibinfo{author}{\bibfnamefont{Weinberg}, \bibnamefont{S.}}:
  \emph{\bibinfo{title}{The Quantum Theory of Fields, Vol. II}}
  (\bibinfo{publisher}{Cambridge University Press}, \bibinfo{year}{2001}): p. 437

\bibitem{AFG}
\bibinfo{author}{\bibfnamefont{Arafune}, \bibnamefont{J.}},
  \bibinfo{author}{\bibfnamefont{Freund}, \bibnamefont{P.G.O.}},
  \bibinfo{author}{\bibfnamefont{Goebel}, \bibnamefont{C.J.}}:
\bibinfo{title}{Topology of Higgs fields}.
\bibinfo{journal}{J. Math. Phys.}
  \textbf{\bibinfo{volume}{16}}, \bibinfo{pages}{433-437} (\bibinfo{year}{1975})

\bibitem{BOG}
\bibinfo{author}{\bibfnamefont{Bogomol'nyi}, \bibnamefont{E.B.}}:
\bibinfo{title}{The stability of classical solutions}.
  \bibinfo{journal}{Sov. J. Nucl. Phys.} \textbf{\bibinfo{volume}{24}},
  \bibinfo{pages}{449-454} (\bibinfo{year}{1976}) [Yad. Fiz. \textbf{24}, 861 (1976)]

\bibitem{STON1}
\bibinfo{author}{\bibfnamefont{Singleton}, \bibnamefont{D.}}:
\bibinfo{title}{Does magnetic charge imply a massive photon?}
  \bibinfo{journal}{Int. J. Theor. Phys.} \textbf{\bibinfo{volume}{35}},
  \bibinfo{pages}{2419-2426} (\bibinfo{year}{1996})

\bibitem{RIPKA}
\bibinfo{author}{\bibfnamefont{Ripka}, \bibnamefont{G.}}:
  \emph{\bibinfo{title}{Dual Superconductor Models of Color Confinement}}
  (\bibinfo{publisher}{Springer, Berlin}, \bibinfo{year}{2005}) [arXiv: hep-ph/0310102]

\bibitem{GREENSITE}
\bibinfo{author}{\bibfnamefont{Greensite}, \bibnamefont{J.}}:
\bibinfo{title}{Some current approaches to the confinement problem}.
  \bibinfo{journal}{Acta Phys. Pol. B} \textbf{\bibinfo{volume}{40}},
  \bibinfo{pages}{3355-3407} (\bibinfo{year}{2009})

\bibitem{HEM1}
\bibinfo{author}{\bibfnamefont{Nandan}, \bibnamefont{H.}},
\bibinfo{author}{\bibfnamefont{Anna}, \bibnamefont{T.}},
\bibinfo{author}{\bibfnamefont{Chandola}, \bibnamefont{H.C.}}:
\bibinfo{title}{Dyon condensation and colour confinement in dual QCD}.
  \bibinfo{journal}{Europhys. Lett.} \textbf{\bibinfo{volume}{67}},
  \bibinfo{pages}{746-752} (\bibinfo{year}{2004})

\bibitem{HEM2}
\bibinfo{author}{\bibfnamefont{Nandan}, \bibnamefont{H.}},
\bibinfo{author}{\bibfnamefont{Chandola}, \bibnamefont{H.C.}},
\bibinfo{author}{\bibfnamefont{Dehnen}, \bibnamefont{H.}}:
\bibinfo{title}{Magnetic symmetry, Regge trajectories, and the linear confinement in dual QCD}.
  \bibinfo{journal}{Int. J. Theor. Phys.} \textbf{\bibinfo{volume}{44}},
  \bibinfo{pages}{457-469} (\bibinfo{year}{2005})


\end{thebibliography}
\end{document}